\documentclass{PoS}

\usepackage{graphicx}
\usepackage{psfrag}
\usepackage{subfigure}

\usepackage{tabulary}
\newlength{\figwidth}
\title{Infrared exponents of gluon and ghost propagators from Lattice QCD}

\ShortTitle{Infrared exponents of gluon and ghost propagators from Lattice QCD}

\author{Orlando Oliveira and \speaker{Paulo J. Silva}\\
        Centro de F\'{\i}sica Computacional,
        Departamento de F\'{\i}sica,
        Universidade de Coimbra, P-3004-516 Coimbra,
        Portugal\\
        E-mail: \email{psilva@teor.fis.uc.pt},
        \email{orlando@teor.fis.uc.pt}}

\abstract{The compatibility of the pure power law infrared solution of QCD Dyson-Schwinger equations (DSE) and lattice data for the gluon and ghost propagators in Landau gauge is discussed. For the gluon propagator, the lattice data is compatible with the DSE infrared solution with an exponent $\kappa\sim0.53$, measured using a technique that suppresses f\mbox{}inite volume effects and allows to model these corrections to the lattice data. For the ghost propagator, the lattice data does not seem to follow the infrared DSE power law solution.}

\FullConference{The XXV International Symposium on Lattice Field Theory\\
		 July 30 - August 4 2007\\
		 Regensburg, Germany}

\begin{document}

The problem of quark and gluon conf\mbox{}inement in QCD is still not understood \cite{alk_green}. In the Landau gauge,
two proposed confinement mechanisms, the Kugo-Ojima and the Gribov-Zwanziger scenarios, relate gluon conf\mbox{}inement to the infrared properties of gluon and ghost propagators.

The investigation of the infrared behaviour of gluon and ghost propagators requires a non-perturbative formulation of QCD such as the Dyson-Schwinger equations (DSE) or the lattice formulation of QCD. The nature of the lattice QCD approach allows for a study of the propagators including all non-perturbative physics. However, the lattice is limited to a f\mbox{}inite volume, therefore the access to the deep infrared regime of QCD requires a proper analysis of finite volume effects.  On the other hand, the DSE (for a recent review see \cite{Fisch05}) were solved analitically in the infrared, predicting a pure power law behaviour for the gluon and ghost dressing functions,
\begin{equation}
Z_{gluon}(q^2)\equiv q^2 D( q^2 )\sim(q^2)^{2\kappa}\,\,\,\,\,,\,\,\,\,\,\,Z_{ghost}(q^2)\equiv q^2 G( q^2 )\sim(q^2)^{-\kappa} .
\label{ppl}
\end{equation}
The solution requires a truncation of an inf\mbox{}inite tower of equations and a parametrization of a number of vertices. Assuming ghost dominance and a bare ghost-gluon vertex, DSE estimated an exponent
 $\kappa=0.595$ \cite{lerche}. This value for $\kappa$, being above 0.5, implies a vanishing gluon propagator and a divergent ghost dressing function, in agreement with the conf\mbox{}inement criteria referred above. Other studies \cite{pawl, gies, tisq, llanes, fischer-pawl} gave further support to this picture.

On the lattice, in order to test reliably for the power law solutions  (\ref{ppl}) one needs very large 
volumes. A possible cheaper solution could be the use of large asymmetric lattices, i.e. lattices
as $L_s^3\times L_t$, with a large $L_t$ --- see \cite{asymmetric} and references there in. Although there are non-negligible f\mbox{}inite volume effects due to the small spatial lattice extension, the large temporal lattice extension allow the access to momenta below 100-200 MeV, and a direct
test on the validity of the solution (\ref{ppl}).

So far, we have computed $\kappa$ by a direct f\mbox{}it of (\ref{ppl}) to the asymmetric lattice data. The
results show that $\kappa$ increases with the spatial lattice volume \cite{asymmetric}. In this proceeding we report on a technique \cite{ratios} that seems to provide a volume independent value for the gluon exponent $\kappa$.

\begin{table}[b]
\begin{center}
\begin{tabular}{|r|@{\hspace{0.2cm}}r|
                    @{\hspace{0.2cm}}r|
                    @{\hspace{0.2cm}}r|
                    @{\hspace{0.2cm}}r|}
\hline
   Lattice           & Update
                     & Therm.
                     & Sep.
                     & \# Conf. \\
\hline
   $8^3 \times 256$   & 7OVR+4HB & 1500 & 1000 & 80  \\
   $10^3 \times 256$  & 7OVR+4HB & 1500 & 1000 & 80  \\
   $12^3 \times 256$  & 7OVR+4HB & 1500 & 1000 & 80  \\
   $14^3 \times 256$  & 7OVR+4HB & 3000 & 1000 & 128  \\
   $16^3 \times 256$  & 7OVR+4HB & 3000 & 1500 & 155 \\  
   $18^3 \times 256$  & 7OVR+4HB & 2000 & 1000 & 150  \\
\hline
   $16^3 \times 128$  & 7OVR+2HB & 3000 & 3000 & 164  \\
\hline
\end{tabular}
\caption{Lattice setup. All sets of conf\mbox{}igurations were generated
using a combined Monte Carlo sweep of overrelaxation (OVR) and heat
bath (HB) updates. The number of thermalization (Therm.)
and separation (Sep.) sweeps refers to
combined sweeps.} \label{Uvol}
\end{center}
\end{table}

The lattice setup used in this work is summarized in table \ref{Uvol}. The conf\mbox{}igurations were gauge f\mbox{}ixed to the Landau gauge using a Fourier accelerated Steepest Descent method, starting from the identity gauge transformation. Then, the gluon propagator was computed using the same def\mbox{}initions as in \cite{asymmetric}. The ghost propagator was computed using a plane-wave source \cite{cucc97}. Statistical errors on the propagators were computed using the jackknife method. Otherwise, the statistical errors were computed using the bootstrap method with a 68\% conf\mbox{}idence level.

On the lattice, the f\mbox{}inite volume and discretization effects can be suppressed by def\mbox{}ining ratios between similar quantities.  For example,  consider some quantity $A(x)$. Suppose that the lattice effects are given by $1+\delta(x)$, with $\delta \ll 1$ . If $x'\sim x$, one can write 
\begin{equation}
\frac{A(x')(1+\delta')}{A(x)(1+\delta)} \simeq \frac{A(x')}{A(x)}(1+\delta')(1-\delta)  \sim  \frac{A(x')}{A(x)}(1-\delta^2) ,
\end{equation}
i.e. the error on the ratio is of second order in $\delta$. 

\begin{figure}[t]
\vspace*{0.5cm}
\begin{center}
\psfrag{EIXOX}{{\small $R_{q}[n]$}}
\psfrag{EIXOY}{{\small $R_{Z}^{gluon}[n]$}}
\includegraphics[width=\figwidth]{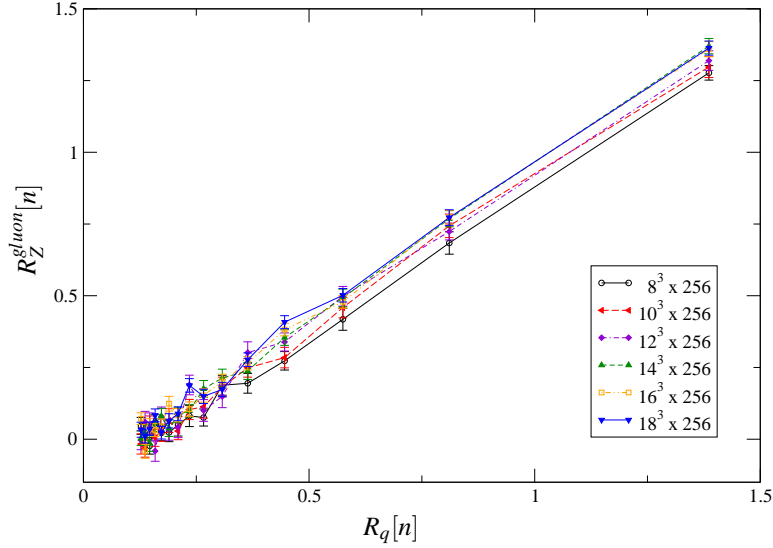}
\caption{$R_{Z}^{gluon}[n]$ as a function of  $R_{q}[n]$ for the lattices with $L_t=256$.}
\label{glueratios}
\end{center}
\end{figure}

Let's take the gluon propagator. If, in the continuum limit, we have $Z(q^2)=\omega (q^2)^{2\kappa}$, on a f\mbox{}inite lattice we have, in general,
\begin{equation}
   Z_{latt}(q^2)=\omega (q^2)^{2\kappa}\Delta(q)
\label{ppl:latt}
\end{equation}
where $\Delta(q)$ can be viewed as a multiplicative correction to the continuum function. Then,
taking ratios of the gluon dressing function between consecutive temporal momenta,
\begin{equation}
q[n]=q_4[n]=\frac{2}{a} ~ \sin \Big( \frac{\pi n}{L_t} \Big)
 \, ,
  \hspace{0.5cm}  n ~ = ~ 0, \, 1, \, \dots \, \frac{L_t}{2} \, 
\end{equation}
and taking logarithms, one gets
\begin{equation}
\ln \left[ \frac{Z_{latt}(q^2[n+1])}{Z_{latt}(q^2[n])} \right] ~  = ~  
2  \kappa \ln \left[ \frac{q^2[n+1]}{q^2[n]} \right] ~ + ~ C(q).
\label{glue.rat.1}
\end{equation}
Def\mbox{}ining
\begin{equation}
R_{Z}^{gluon}[n] \equiv  \ln \left[ \frac{Z_{latt}(q^2[n+1])}{Z_{latt}(q^2[n])} \right]\,\,\,\, ,\,\,\,\,
R_{q}[n] \equiv  \ln \left[\frac{q^2[n+1]}{q^2[n]} \right],
\end{equation}
equation (\ref{glue.rat.1}) becomes
\begin{equation}
R_{Z}^{gluon}[n] ~  = ~  
2  \kappa R_{q}[n] ~ + ~ C(q).
\label{eq_ratio}
\end{equation}

Figure \ref{glueratios} shows these functions for the  asymmetric lattices with $L_t=256$. For each lattice size, the data points def\mbox{}ine straight lines in the infrared region, i.e. it
seems that $C(q)$ is a constant, $C(q) \equiv C$. This hipothesis can be tested by f\mbox{}itting 
$R_{Z}^{gluon}[n]$ to a linear function of $R_{q}[n]$. The corresponding values for  $\kappa$ and $C$ are reported in table \ref{linear_glue_ratio}. Note that the measured $\kappa$ values are stable against variation of the f\mbox{}itting range and spatial lattice size. Furthermore, within one standard deviation,
$\kappa>0.5$ as in the solution of the DSE.

\begin{table}[t]
\begin{center}
\footnotesize
\begin{tabular}{|l|l|l|l|l|l|l|}
\hline
    $L_s$    & $q_{max}$:  &  \multicolumn{1}{c|}{\hspace*{0.3cm} 191 MeV \hspace*{0.3cm} } 
               &  \multicolumn{1}{c|}{\hspace*{0.3cm} 238 MeV \hspace*{0.3cm} }
               &  \multicolumn{1}{c|}{\hspace*{0.3cm} 286 MeV \hspace*{0.3cm} }
               &  \multicolumn{1}{c|}{\hspace*{0.3cm} 333 MeV \hspace*{0.3cm} }
               &  \multicolumn{1}{c|}{\hspace*{0.3cm} 381 MeV \hspace*{0.3cm} }   \\
\hline
8  & $\kappa$  & $0.526(27)$  & $0.531(19)$  & $0.531(13)$ & $0.522(16)$  & $0.527(12)$ \\
   & $C$ & $-0.179(54)$      & $-0.194(34)$      & $-0.193(19)$  & $-0.171(28)$   & $-0.184(18)$  \\
   & $\chi^2/d.o.f. $ & \multicolumn{1}{r|}{ $0.12$}     & \multicolumn{1}{r|}{ $0.11$}   &\multicolumn{1}{r|}{ $0.08$ }  & \multicolumn{1}{r|}{ $0.48$}   &\multicolumn{1}{r|}{ $0.54$ }  \\
\hline
10  & $\kappa$  & $0.511(35)$  & $0.531(25)$  & $0.525(21)$ & $0.523(17)$  & $0.527(16)$ \\
%   & $C$ & $-0.114(66)$      & $-0.161(42)$      & $-0.146(30)$  & $-0.144(21)$   & $-0.150(19)$  \\
   & $\chi^2/d.o.f. $ & \multicolumn{1}{r|}{ $0.69$}     & \multicolumn{1}{r|}{ $0.98$}   &\multicolumn{1}{r|}{ $0.74$ }  & \multicolumn{1}{r|}{ $0.56$}   &\multicolumn{1}{r|}{ $0.50$ }  \\
\hline
12  & $\kappa$  & $0.509(31)$  & $0.517(21)$  & $0.508(18)$ & $0.521(18)$  & $0.530(14)$ \\
   & $C$ & $-0.094(56)$      & $-0.112(35)$      & $-0.094(25)$  & $-0.119(27)$   & $-0.138(18)$  \\
   & $\chi^2/d.o.f. $ & \multicolumn{1}{r|}{ $0.11$}     & \multicolumn{1}{r|}{ $0.16$}   &\multicolumn{1}{r|}{ $0.33$ }  & \multicolumn{1}{r|}{ $0.84$}   &\multicolumn{1}{r|}{ $1.03$ }  \\
\hline
14  & $\kappa$  & $0.536(24)$  & $0.540(19)$  & $0.548(16)$ & $0.545(12)$  & $0.542(11)$ \\
%   & $C$ & $-0.114(44)$      & $-0.123(30)$      & $-0.140(21)$  & $-0.134(15)$   & $-0.127(12)$  \\
   & $\chi^2/d.o.f. $ & \multicolumn{1}{r|}{ $0.33$}     & \multicolumn{1}{r|}{ $0.20$}   &\multicolumn{1}{r|}{ $0.39$ }  & \multicolumn{1}{r|}{ $0.34$}   &\multicolumn{1}{r|}{ $0.34$ }  \\
\hline
16  & $\kappa$  & $0.539(22)$  & $0.528(17)$  & $0.534(12)$ & $0.536(12)$  & $0.539(11)$ \\
   & $C$ & $-0.125(43)$      & $-0.102(30)$      & $-0.112(19)$  & $-0.118(14)$   & $-0.123(12)$  \\
   & $\chi^2/d.o.f. $ & \multicolumn{1}{r|}{ $1.77$}     & \multicolumn{1}{r|}{ $1.24$}   &\multicolumn{1}{r|}{ $0.96$ }  & \multicolumn{1}{r|}{ $0.78$}   &\multicolumn{1}{r|}{ $0.68$ }  \\
\hline
18  & $\kappa$  & $0.529(20)$  & $0.516(16)$  & $0.523(14)$ & $0.536(11)$  & $0.5398(95)$ \\
   & $C$ & $-0.099(36)$      & $-0.068(25)$      & $-0.085(19)$  & $-0.111(14)$   & $-0.119(13)$  \\
   & $\chi^2/d.o.f. $ & \multicolumn{1}{r|}{ $0.39$}     & \multicolumn{1}{r|}{ $0.77$}   &\multicolumn{1}{r|}{ $0.85$ }  & \multicolumn{1}{r|}{ $1.79$}   &\multicolumn{1}{r|}{ $1.58$ }  \\
\hline
\end{tabular}
\caption{Linear f\mbox{}its of $R_{Z}^{gluon}[n]$ as a function of $R_{q}[n]$.}
\label{linear_glue_ratio}
\end{center}
\end{table}

The results of the linear f\mbox{}it suggest a parametrization of the f\mbox{}inite volume effects. From the def\mbox{}inition of
$\Delta$, it follows that $\Delta (q[n+1]) =  \Delta (q[n]) e^C$ and
\begin{equation}
  \frac{d \Delta (q)}{d q} \sim 
         \frac{\Delta (q[n+1]) - \Delta(q[n])}{q[n+1] - q[n]} 
    \sim \Delta (q) \, \frac{e^C - 1 }{\frac{2 \pi}{a L_t}} 
    ~ = ~ \Delta (q) \, A
\end{equation}
where A is a constant. The integration of this equation gives
\begin{equation}
  \Delta (q) ~ = ~ \Delta_0 \, e^{A q} \,,
\end{equation}
i.e. the lattice dressing function is given by
\begin{equation}
  Z_{Lat} (q^2) ~ = ~ \omega \left( q^2 \right)^{2 \kappa} \, e^{A q}  \,;
  \label{ZCorrLat}
\end{equation}
note that now $\omega$ has absorved the constant $\Delta_0$. The 
f\mbox{}inite volume effects are summarized by the constant $A$. The results of f\mbox{}itting the lattice data to
(\ref{ZCorrLat}) are reported in table \ref{fit_glue_exp}. The $\kappa$ values in tables
 \ref{linear_glue_ratio} and \ref{fit_glue_exp} are, as expected, essentially the same. Furthermore,
the constants $A$ and $C$ are, in general, decreasing functions of the (spatial) lattice volume, and should go to zero in the inf\mbox{}inite volume limit.

\begin{table}[t]
\begin{center}
\footnotesize
\begin{tabular}{|l|l|l|l|l|l|l|}

\hline
       $L_s$    & $q_{max}$:   &  \multicolumn{1}{c|}{\hspace*{0.3cm} 191 MeV \hspace*{0.3cm} } 
               &  \multicolumn{1}{c|}{\hspace*{0.3cm} 238 MeV \hspace*{0.3cm} }
               &  \multicolumn{1}{c|}{\hspace*{0.3cm} 286 MeV \hspace*{0.3cm} }
               &  \multicolumn{1}{c|}{\hspace*{0.3cm} 333 MeV \hspace*{0.3cm} }
               &  \multicolumn{1}{c|}{\hspace*{0.3cm} 381 MeV \hspace*{0.3cm} }   \\
\hline
8  & $\kappa$  & $0.526(26)$  & $0.533(19)$  & $0.534(11)$ & $0.523(10)$  & $0.524(9)$ \\
   & $A(GeV^{-1})$ & $-3.75\pm1.1$ & $-4.06(68)$      & $-4.11(34)$  & $-3.69(28)$   & $-3.73(23)$  \\
   & $\chi^2/d.o.f. $ & \multicolumn{1}{r|}{ $0.09$}     & \multicolumn{1}{r|}{ $0.12$}   &\multicolumn{1}{r|}{ $0.08$ }  & \multicolumn{1}{r|}{ $0.62$}   &\multicolumn{1}{r|}{ $0.51$ }  \\
\hline
10  & $\kappa$  & $0.511(27)$  & $0.536(22)$  & $0.534(17)$ & $0.531(14)$  & $0.534(13)$ \\
%   & $A(GeV^{-1})$ & $-2.3\pm1.1$ & $-3.40(69)$  & $-3.33(51)$  & $-3.22(37)$  & $-3.30(29)$  \\
   & $\chi^2/d.o.f. $ & \multicolumn{1}{r|}{ $0.53$}     & \multicolumn{1}{r|}{ $1.08$}   &\multicolumn{1}{r|}{ $0.73$ }  & \multicolumn{1}{r|}{ $0.58$}   &\multicolumn{1}{r|}{ $0.49$ }  \\
\hline
12  & $\kappa$  & $0.508(31)$  & $0.515(22)$  & $0.507(15)$ & $0.520(12)$  & $0.537(9)$ \\
   & $A(GeV^{-1})$ & $-1.9\pm1.2$      & $-2.25(78)$      & $-1.92(46)$  & $-2.40(36)$   & $-2.96(23)$  \\
   & $\chi^2/d.o.f. $ & \multicolumn{1}{r|}{ $0.07$}     & \multicolumn{1}{r|}{ $0.12$}   &\multicolumn{1}{r|}{ $0.24$ }  & \multicolumn{1}{r|}{ $0.84$}   &\multicolumn{1}{r|}{ $1.94$ }  \\
\hline
14  & $\kappa$  & $0.538(23)$  & $0.542(18)$  & $0.552(14)$ & $0.551(11)$  & $0.546(9)$ \\
%   & $A(GeV^{-1})$ & $-2.42(87)$ & $-2.62(59)$      & $-3.00(41)$  & $-2.96(29)$   & $-2.80(21)$  \\
   & $\chi^2/d.o.f. $ & \multicolumn{1}{r|}{ $0.24$}     & \multicolumn{1}{r|}{ $0.17$}   &\multicolumn{1}{r|}{ $0.47$ }  & \multicolumn{1}{r|}{ $0.36$}   &\multicolumn{1}{r|}{ $0.45$ }  \\
\hline
16  & $\kappa$  & $0.541(22)$  & $0.532(16)$  & $0.535(10)$ & $0.539(9)$  & $0.543(8)$ \\
   & $A(GeV^{-1})$ & $-2.67(84)$ & $-2.29(54)$  & $-2.39(31)$  & $-2.53(24)$   & $-2.66(18)$  \\
   & $\chi^2/d.o.f. $ & \multicolumn{1}{r|}{ $1.15$}     & \multicolumn{1}{r|}{ $0.78$}   &\multicolumn{1}{r|}{ $0.55$ }  & \multicolumn{1}{r|}{ $0.50$}   &\multicolumn{1}{r|}{ $0.54$ }  \\
\hline
18  & $\kappa$  & $0.529(20)$  & $0.516(15)$  & $0.523(12)$ & $0.539(9)$  & $0.550(8)$ \\
   & $A(GeV^{-1})$ & $-2.05(79)$  & $-1.50(51)$ & $-1.75(33)$  & $-2.31(24)$   & $-2.66(20)$  \\
   & $\chi^2/d.o.f. $ & \multicolumn{1}{r|}{ $0.28$}  & \multicolumn{1}{r|}{ $0.59$}   &\multicolumn{1}{r|}{ $0.54$ }  & \multicolumn{1}{r|}{ $2.14$}   &\multicolumn{1}{r|}{ $2.71$ }  \\
\hline
\end{tabular}
\caption{Fits of the gluon dressing function to the pure power law with exponential correction.}
\label{fit_glue_exp}
\end{center}
\end{table}

The ratio method provides similar results when used with the $16^3\times128$ lattice data. As shown
in \cite{asymmetric}, the gluon data from $16^3\times128$ and $16^3\times256$ lattices are compatible within errors and one expects similar values for the constant $A$ for these two lattices. Indeed,
the f\mbox{}its to (\ref{ZCorrLat}), see  table \ref{urat128}, give essentially the same $A$. Given the relation between $A$ and $C$, 
\begin{equation}
   A = \frac{e^C-1}{\frac{2\pi}{aL_t}}\sim C \frac{aL_t}{2\pi},
\end{equation}
we expect, as observed, that  $C_{128}\simeq 2\times C_{256}$. Furthermore, from table \ref{urat128},
one can conclude that the effects of Gribov copies are not resolved by the statistical precision of our simulation.

\begin{table}[b]
\begin{center}
\footnotesize
\begin{tabular}{lcccccc}
\hline
       & \multicolumn{3}{c}{Ratios} & \multicolumn{3}{c}{Modelling} \\
\hline
$16^3\times L_t$ & $\kappa$ & $C$ & $\chi^2/dof$ & $\kappa$ & $A(GeV^{-1})$ &$\chi^2/dof$ \\
\hline
$L_t=256$ & $0.539(11)$ & $-0.123(12)$ & $0.68$ & $0.543(8)$ & $-2.66(18)$ & $0.54$ \\
$L_t=128$ \small{[ID]} & $0.541(19)$ & $-0.239(38)$ & $0.01$ & $0.542(20)$ & $-2.56(39)$ & $0.01$ \\
$L_t=128$ \small{[CEASD]} & $0.539(19)$ & $-0.234(36)$ & $0.15$ & $0.539(18)$ & $-2.47(36)$ & $0.10$ \\
\hline
\end{tabular}
\caption{Results obtained for the lattices with $L_s=16$ ($q<381$MeV). For the lattice $16^3\times128$, two gauge f\mbox{}ixing methods were considered. ID stands for a gauge f\mbox{}ixing started from the identity gauge transformation, and CEASD stands for the gauge f\mbox{}ixing method devised in \cite{cpc2004}, aiming to f\mbox{}ind the global maximum of $F_U[g]$.  }
\label{urat128}
\end{center}
\end{table}

\begin{figure}[t]
\vspace*{0.5cm}
\begin{center}
\psfrag{EIXOX}{{ $x\equiv1/V$}}
\psfrag{EIXOY}{{ $D(0)$}}
\includegraphics[width=\figwidth]{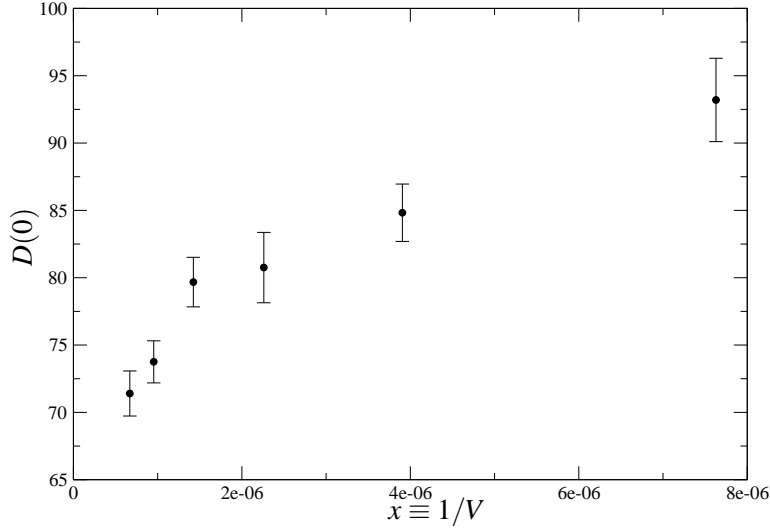}
\caption{Values of $D(0)$ as a function of $1/V$.}
\label{D_0}
\end{center}
\end{figure}

Note that the estimated gluon infrared exponent $\kappa\sim0.53$ implies a vanishing gluon propagator at zero momentum. Figure \ref{D_0} shows the bare $D(0)$ as a function of $x\equiv1/V$. In what concerns the inf\mbox{}inite volume $D(0)$, a linear ($D_{\infty}(0)+bx$) or quadratic ($D_{\infty}(0)+bx+cx^2$) extrapolations give a non-zero $D_{\infty}(0)$, but a power law ($ax^b$) extrapolation, which implies $D_{\infty}(0)\equiv 0$, is also possible, and gives $b=0.10$. Curiously, this value is very close to the f\mbox{}igure reported in a recent investigation of the DSE on a torus \cite{dsetorus2}, $b\sim 0.095$.

In what concerns the ghost propagator, see \cite{ratios} for details, we were not able to see a linear behaviour of $R_{Z}^{ghost}[n]$ as a function of  $R_{q}[n]$. Assuming that the ghost dressing function really follows a pure power law in the infrared, the only result one could extract is a kind of lower bound for the ghost infrared exponent, $\kappa\sim0.29$.

\begin{figure}[t]
\vspace*{0.5cm}
\begin{center}
\psfrag{EIXOX}{{ $q (GeV)$ }}
\psfrag{EIXOY}{{ $\alpha_S(q^2)$}}
\includegraphics[width=\figwidth]{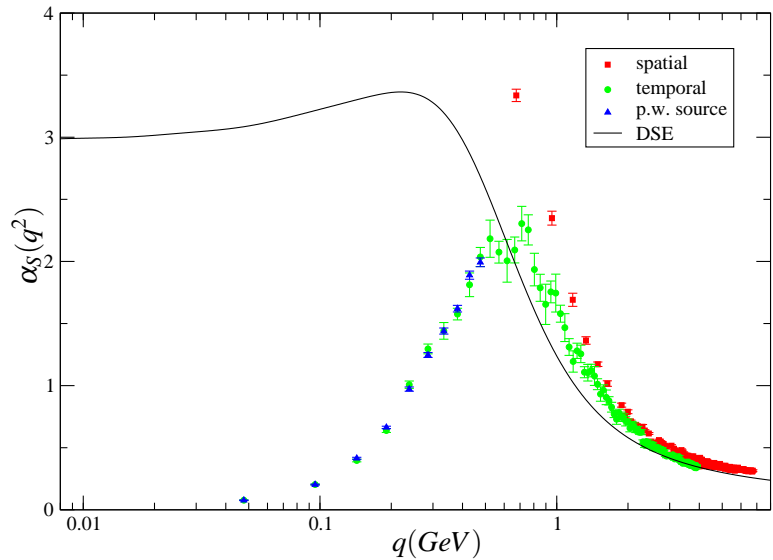}
\end{center}
\caption{Comparison of results for $\alpha_S(q^2)$ using both lattice QCD and DSE.}
\label{alpha}
\end{figure}

Finally, a few words about the running coupling def\mbox{}ined from the gluon and ghost propagators,
\begin{equation}
  \alpha_S ( q^2 ) ~ = ~ \alpha_S ( \mu^2 ) \, Z^2_{ghost} ( q^2 ) \,
                                               Z_{gluon} ( q^2 ) \, .
\label{alfaS}
\end{equation}
For the lattice data strong coupling constant see f\mbox{}igure \ref{alpha}.  If the DSE  predict a f\mbox{}inite non-zero value for $\alpha_S(0)= 2.972$, the lattice data seems to go to zero. This behaviour is in agreement with the solution of the Dyson-Schwinger equations on a torus \cite{dsetorus2}. According to this study, one should go to even larger lattices to become closer to the continuum. This is a very ambitious challenge for the next years.

\acknowledgments

This work was supported by FCT via grant SFRH/BD/10740/2002, and partially supported by projects POCI/FP/63436/2005 and POCI/FP/63923/2005.

\end{document}